\documentclass[journal,onecolumn,10pt]{IEEEtran}
\usepackage{amsmath,amssymb,amsthm}
\usepackage{bm}
\usepackage{cite}
\usepackage[font=footnotesize,labelsep=period]{caption}

\allowdisplaybreaks

\newtheoremstyle{hy}{6pt}{6pt}{\normalfont}{}{\bfseries}{.}{.5em}{}
\theoremstyle{hy}
\newtheorem{theorem}{Theorem}
\newtheorem{lemma}{Lemma}
\newtheorem{corollary}{Corollary}

\newtheorem{definitionx}{Definition}
\newtheorem{remarkx}{Remark}
\newenvironment{definition}{\begin{definitionx}}{\hfill$\Box$\end{definitionx}}
\newenvironment{remark}{\begin{remarkx}}{\hfill$\Box$\end{remarkx}}
\renewenvironment{proof}[1][Proof]{\par\noindent\textit{#1:}\ \ }{\hfill$\Box$\par\medskip}

\newcommand{\bX}{\bm{X}}
\newcommand{\bZ}{\bm{Z}}
\newcommand{\bW}{\bm{W}}
\newcommand{\bx}{\bm{x}}
\newcommand{\bz}{\bm{z}}
\newcommand{\bu}{\bm{u}}
\newcommand{\bj}{\bm{j}}
\newcommand{\bJ}{\bm{J}}
\newcommand{\bR}{\bm{R}}
\newcommand{\bP}{\bm{P}}
\newcommand{\bGamma}{\boldsymbol{\Gamma}}
\newcommand{\R}{\mathbb{R}}
\newcommand{\E}{\mathbb{E}}

\newcommand{\cG}{\mathcal{G}}
\newcommand{\cS}{\mathcal{S}}
\newcommand{\cA}{\mathcal{A}}
\newcommand{\cX}{\mathcal{X}}
\newcommand{\cZ}{\mathcal{Z}}
\newcommand{\cN}{\mathcal{N}}
\newcommand{\cD}{\mathcal{D}}
\newcommand{\cE}{\mathcal{E}}
\newcommand{\cK}{\mathcal{K}}
\newcommand{\cR}{\mathcal{R}}
\newcommand{\cU}{\mathcal{U}}

\newcommand{\cT}{\mathcal{T}}
\newcommand{\cM}{\mathcal{M}}
\newcommand{\sfC}{\mathsf{C}}
\newcommand{\tr}{\operatorname{tr}}
\newcommand{\plimi}{\mathop{\mbox{p-}\liminf}_{n\to\infty}}
\newcommand{\plims}{\mathop{\mbox{p-}\limsup}_{n\to\infty}}
\newcommand{\st}{\,\big|\,}
\newcommand{\Iinf}{\underline{I}}
\newcommand{\Isup}{\overline{I}}
\newcommand{\Sc}{{\cS^c}}

\begin{document}

\title{Information-Spectrum Methods for Multiple-Access\\ Channels---Part~III: The Strong Converse\\ in the $K$-User Gaussian Case}
\author{Te Sun Han, \textit{Life Fellow, IEEE} and Hideki Yagi, \textit{Member, IEEE}%
\thanks{T. S. Han is with The University of Electro-Communications, Chofu, Tokyo 182-8585, Japan (email: han@is.uec.ac.jp).}%
\thanks{H. Yagi is with The University of Electro-Communications, Chofu, Tokyo 182-8585, Japan (email: h.yagi@uec.ac.jp).}}
\maketitle

\begin{abstract}
This paper, Part III of a series, treats $K$-user multiple-access channels (MACs). Its starting point is a pair of fundamental
lemmas, a random-coding lemma of Feinstein type and a converse lemma of Verd\'u--Han type, which are the $K$-user extensions of the
two lemmas underlying Han's information spectrum approach to the general two-user MAC. From them we derive the capacity region of
the general $K$-user MAC with cost constraint and show that it contains the polymatroid capacity region of the $K$-user Gaussian MAC.
We then prove, by an elementary information spectrum calculation that adapts a diffuse--exceptional subspace decomposition used in
recent second-order converses for the Gaussian MAC, that for every subset $\cS$ of users and every independent input satisfying the power constraints, the conditional information
spectral-sup rate of the $K$-user Gaussian MAC does not exceed $\frac12\log(1+P_\cS/N)$, where $P_\cS$ is the total power of the users
in $\cS$, with a probability bound that does not depend on the input. Consequently, the strong converse for the $K$-user Gaussian MAC,
in the sense that the error probability tends to one for every rate tuple outside the capacity region, follows directly from the
converse lemma, without the wringing technique and without hypothesis-testing meta-converses. This strong converse was first established by Fong and Tan, whose per-blocklength bounds yield it, through wringing and quantization. It is the key
ingredient in the converse parts of the coding theorems for mixed Gaussian MACs in Part I, where the $\varepsilon$-strong converse
would not suffice.
\end{abstract}

\begin{center}
\textbf{Index Terms}
\end{center}
\begin{quote}\small
Information spectrum, multiple-access channel, $K$-user Gaussian MAC, capacity region, polymatroid, strong converse,
$\varepsilon$-strong converse, mixed channel
\end{quote}

%% ====================================================================
\section{Introduction}\label{sec:intro}

\subsection{Background}

In Part I \cite{HY-I} and Part II \cite{HY-II} of this series, the $\varepsilon$-capacity regions of mixed multiple-access
channels (MACs) were studied, for Gaussian and quasi-static fading components in Part I and for finite-alphabet
memoryless components in Part II, with one and the same general formula \cite[Theorem~1]{HY-I}, \cite[Theorem~1]{HY-II},
a slight modification of the information spectrum formula of Han \cite[Theorem~5]{Han98}, as the starting point of the direct parts. On the converse side, the
mixed Gaussian MACs of Part I required one more ingredient: the \emph{strong converse property} of the Gaussian MAC
\cite[Definition~2 and Theorem~3]{HY-I}, i.e., the property that the average error probability of any sequence of codes
whose rate tuple lies outside the capacity region tends to one. This property, credited in Part I to Fong and Tan
\cite{FT16} \cite[Remark~5]{HY-I}, is applied there to each component channel separately, and combined with the identity
expressing the error probability over a mixed MAC as the average of the error probabilities over its components
\cite[Eq.~(28)]{HY-I} it yields the converse parts of the coding theorems for quasi-static fading Gaussian MACs
\cite[Theorems~4, 6 and 7]{HY-I}. For $K$ users, its $K$-user version \cite[Remark~4]{HY-I} is needed. Part I announced
that an information spectrum proof of this $K$-user strong converse, not using the wringing technique, would be given in
Part III. The present paper provides it.

The present Part III treats $K$-user MACs throughout, with emphasis on the $K$-user Gaussian MAC, and goes one step further back than Parts I and II. Its starting
point is not the general formula but the two lemmas from which the general formulas of Han \cite{Han98} are themselves
derived: a random-coding lemma of Feinstein type \cite[Lemma~3]{Han98} and a converse lemma of Verd\'u--Han type
\cite[Lemma~4]{Han98}, the multiterminal counterpart of \cite[Theorem~4]{VH94}. We state their natural $K$-user extensions in Sec.~\ref{sec:lemmas} as the \emph{two fundamental
lemmas} (Lemmas~\ref{lem:direct} and \ref{lem:converse}). They are completely general information spectrum lemmas,
valid for arbitrary (nonstationary, nonergodic) MACs with arbitrary alphabets, and they provide a tool that covers the
coding problems for the MAC as a whole: the coding theorems of the present paper, and also the general formulas used as
the starting points of Parts I and II, are derived from them, the remaining work being an evaluation of information spectra
(the information spectrum calculus of \cite[Sec.~I-B]{HY-I}; see \cite[Fig.~1]{HY-I} for an overview of the series).

For a single user ($K=1$), the strong converse for the Gaussian channel goes back to Shannon \cite{Sha59}, and an information
spectrum proof of it was given in \cite[Theorem~3.7.4]{Han03}; second-order refinements were obtained in \cite{Hay09}, \cite{PPV10}.
The present paper extends this information spectrum treatment to $K$ users. The new difficulty is the cross terms between the
inputs of different users, which are independent but otherwise arbitrary.

\subsection{Contributions and Organization}

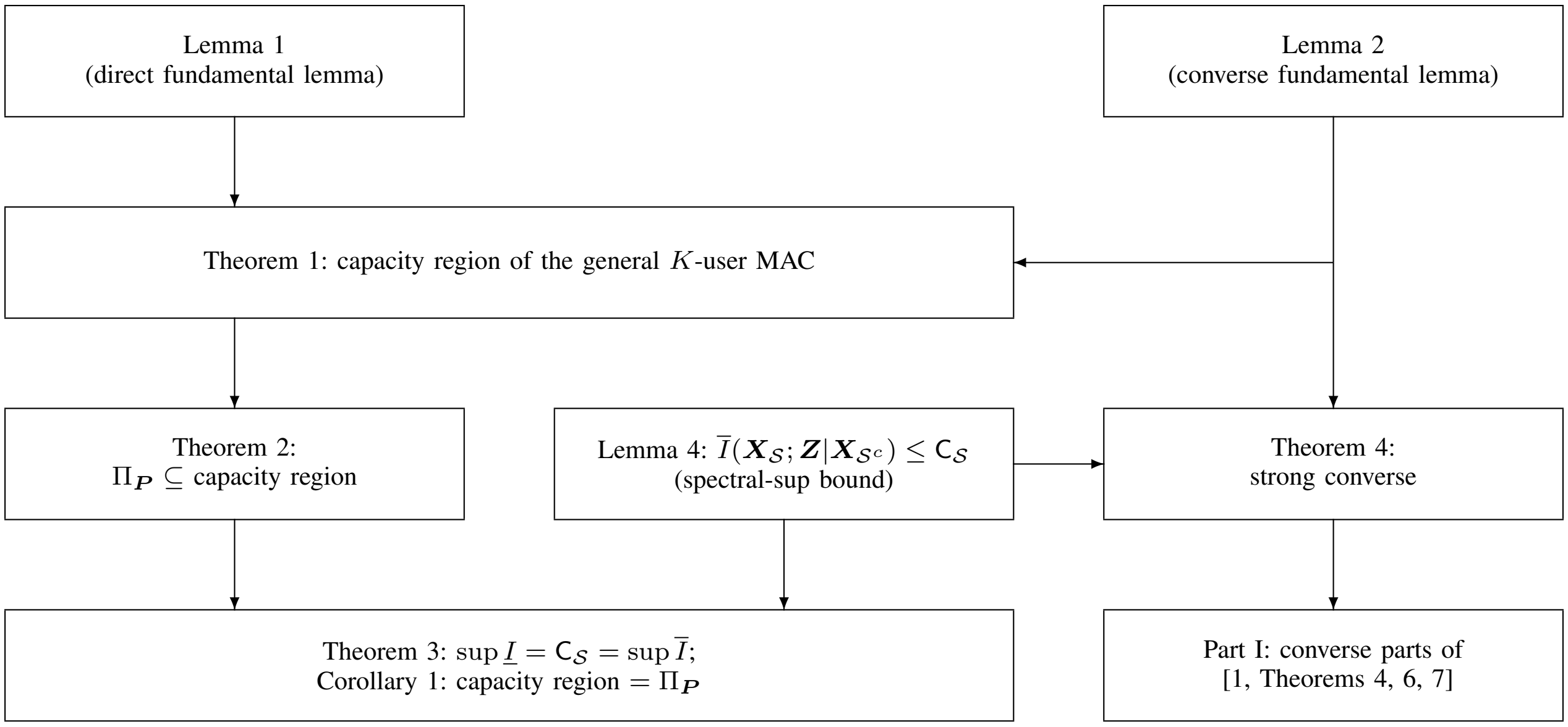
\begin{figure}[t]
\centering
\setlength{\unitlength}{0.0054\textwidth}%
\newcommand{\fbx}[4]{\put(#1,#2){\framebox(#3,12){\parbox{#3\unitlength}{\centering\footnotesize #4}}}}%
\begin{picture}(180,78)(0,22)
\fbx{5}{88}{50}{Lemma~\ref{lem:direct}\\(direct fundamental lemma)}
\fbx{125}{88}{50}{Lemma~\ref{lem:converse}\\(converse fundamental lemma)}
\fbx{5}{66}{110}{Theorem~\ref{thm:general}: capacity region of the general $K$-user MAC}
\fbx{5}{44}{50}{Theorem~\ref{thm:gauss-direct}:\\$\Pi_{\bP}\subseteq$ capacity region}
\fbx{65}{44}{50}{Lemma~\ref{lem:limsup}: $\Isup(\bX_\cS;\bZ|\bX_\Sc)\le\sfC_\cS$\\(spectral-sup bound)}
\fbx{125}{44}{50}{Theorem~\ref{thm:sc}:\\strong converse}
\fbx{5}{22}{110}{Theorem~\ref{thm:sup}: $\sup\Iinf=\sfC_\cS=\sup\Isup$;\\Corollary~\ref{cor:region}: capacity region $=\Pi_{\bP}$}
\fbx{125}{22}{50}{Part~I: converse parts of\\\cite[Theorems~4, 6, 7]{HY-I}}
\put(30,88){\vector(0,-1){10}}
\put(150,88){\line(0,-1){16}}
\put(150,72){\vector(-1,0){35}}
\put(150,72){\vector(0,-1){16}}
\put(30,66){\vector(0,-1){10}}
\put(115,50){\vector(1,0){10}}
\put(30,44){\vector(0,-1){10}}
\put(90,44){\vector(0,-1){10}}
\put(150,44){\vector(0,-1){10}}
\end{picture}
\caption{Logical structure of the paper. The coding theorems are derived from the two fundamental lemmas of Sec.~\ref{sec:lemmas};
Lemma~\ref{lem:limsup} is an evaluation of information spectra and is not derived from them.
The strong converse (Theorem~\ref{thm:sc}) uses only the converse fundamental lemma and the spectral-sup bound of Lemma~\ref{lem:limsup}.
Here $\cS$ is a nonempty subset of the users $[1:K]=\{1,\ldots,K\}$, $\cS^c=[1:K]\setminus\cS$, and $\sfC_\cS=\frac12\log(1+P_\cS/N)$.}
\label{fig:flow}
\end{figure}

The contributions of this paper, all for $K$ users, are as follows (see also Fig.~\ref{fig:flow}).
\begin{itemize}
\item In Sec.~\ref{sec:lemmas} we set up the general $K$-user MAC with cost constraint and state the two fundamental lemmas
(Lemmas~\ref{lem:direct} and \ref{lem:converse}).
\item In Sec.~\ref{sec:general} we derive from the two fundamental lemmas the capacity region of the general $K$-user MAC
with cost constraint in terms of the conditional information spectral-inf rates (Theorem~\ref{thm:general}), the $K$-user
version of \cite[Theorems~1 and 7]{Han98}. The closedness of the union over inputs, which is needed to identify the
capacity region with the union itself, is proved by a diagonal argument (Lemma~\ref{lem:closed}).
\item In Sec.~\ref{sec:gauss-direct} we specialize Theorem~\ref{thm:general} to the stationary memoryless $K$-user Gaussian MAC
and show that the region it describes contains the polymatroid $\Pi_{\bP}$ of the $K$-user Gaussian MAC
(Theorem~\ref{thm:gauss-direct}).
\item In Sec.~\ref{sec:limsup} we prove that, for every nonempty subset $\cS$ of users and every independent input satisfying
the power constraints, the conditional information spectral-sup rate of the $K$-user Gaussian MAC does not exceed
$\sfC_\cS=\frac12\log(1+P_\cS/N)$, with a probability bound that does not depend on the input (Lemma~\ref{lem:limsup}). The proof
is an elementary information spectrum calculation that adapts a diffuse--exceptional subspace decomposition due to Tan
\cite{Tan-cp}, \cite{Tan-sr} (Remark~\ref{rem:tan}). Combining it with
Theorem~\ref{thm:gauss-direct}, we show that the suprema over all admissible inputs of the spectral-inf and spectral-sup
rates both equal $\sfC_\cS$ (Theorem~\ref{thm:sup}), and recover the capacity region $\Pi_{\bP}$ of the $K$-user Gaussian MAC
(Corollary~\ref{cor:region}).
\item In Sec.~\ref{sec:sc} we prove the strong converse for the $K$-user Gaussian MAC in the sense of \cite[Definition~2]{HY-I}
(Theorem~\ref{thm:sc}). The proof uses only the converse fundamental lemma (Lemma~\ref{lem:converse}) and Lemma~\ref{lem:limsup};
neither the wringing technique nor a hypothesis-testing meta-converse is used.
\item In Sec.~\ref{sec:discussion} we discuss the relation to the proof of Fong and Tan \cite{FT16} and to the wringing
arguments of Dueck \cite{Dueck81} and Ahlswede \cite{Ahl82}, and the role of Theorem~\ref{thm:sc} in Part I, for which the
$\varepsilon$-strong converse would not suffice.
\end{itemize}
Sec.~\ref{sec:conc} concludes the paper.

\subsection{How to Read This Paper}

This paper is self-contained and can be read independently of Parts I and II. All definitions used here (the general
$K$-user MAC with cost constraint, the information densities, the $K$-user Gaussian MAC, and the strong converse property) are
stated in Secs.~\ref{sec:lemmas}, \ref{sec:gauss-direct} and \ref{sec:sc}, and all proofs are complete; the references to Parts I and II
serve only to place the results in the series and are not needed to follow any argument.

The main result is the strong converse (Theorem~\ref{thm:sc}). Its proof uses only two ingredients: the converse fundamental lemma
(Lemma~\ref{lem:converse}), a completely general information spectrum lower bound on the error probability, and the spectral-sup bound
(Lemma~\ref{lem:limsup}), an evaluation of the information spectrum of the $K$-user Gaussian MAC. A reader interested only in the strong
converse may therefore read Sec.~\ref{sec:lemmas}, the definition of the Gaussian MAC at the beginning of Sec.~\ref{sec:gauss-direct}, Sec.~\ref{sec:limsup}
up to the proof of Lemma~\ref{lem:limsup}, and Sec.~\ref{sec:sc}. The remaining results (the capacity region of the general $K$-user MAC,
Theorem~\ref{thm:general}; the achievability of the polymatroid, Theorem~\ref{thm:gauss-direct}; and the suprema of the spectral rates,
Theorem~\ref{thm:sup}) are by-products that show how the two fundamental lemmas also yield the coding theorem itself, so that the
whole $K$-user Gaussian MAC is treated from one starting point.

The engineering picture behind Lemma~\ref{lem:limsup} is simple. For a code at a rate above capacity, the converse fundamental lemma says that
the code fails unless the information density exceeds the rate. The information density is, up to a vanishing term, the log-ratio of the
channel to a reference output law; the output energy is at most $n(P_\cS+N)$ by the power constraints, so a Gaussian reference law of that
variance can only be beaten by an input that concentrates its energy in a few directions. Such directions are few, and enlarging the
reference variance there costs a vanishing rate. Hence no input, however correlated across time, can push the information density
above capacity, and every code above capacity fails with probability tending to one.

\subsection{Notation}

Logarithms are natural. For a positive integer $K$ we write $[1:K]\equiv\{1,\ldots,K\}$. For a subset $\cS\subseteq[1:K]$,
$\cS^c\equiv[1:K]\setminus\cS$, and for a tuple $(a_k)_{k\in[1:K]}$ we write $a_\cS\equiv(a_k)_{k\in\cS}$. For real numbers
$R_1,\ldots,R_K$ we write $R_\cS\equiv\sum_{k\in\cS}R_k$, and for positive integers $M_1,\ldots,M_K$ we write
$M_\cS\equiv\prod_{k\in\cS}M_k$; $R_\emptyset=0$ and $M_\emptyset=1$. For a subspace $\cK\subseteq\R^n$, $\Pi_\cK$ denotes the
orthogonal projection onto $\cK$ and $u_\cK\equiv\Pi_\cK u$ for $u\in\R^n$. For a positive semidefinite matrix $A$,
$\lambda_{\max}(A)$ is its largest eigenvalue, and $A\preceq B$ means that $B-A$ is positive semidefinite.
$\cN(\mu,\Sigma)$ is the Gaussian distribution with mean $\mu$ and covariance $\Sigma$.

%% ====================================================================
\section{Two Fundamental Lemmas}\label{sec:lemmas}

\subsection{General $K$-User MAC with Cost Constraint}

Let $\cX_1,\ldots,\cX_K$ be input alphabets and $\cZ$ an output alphabet, all arbitrary (finite, countably infinite,
continuous or abstract). A general $K$-user MAC is a sequence $\bW=\{W^n\}_{n=1}^\infty$ of transition probabilities
$W^n:\cX_1^n\times\cdots\times\cX_K^n\to\cZ^n$, which may be nonstationary and nonergodic. As in \cite[footnote~1]{Han98}
and \cite[Sec.~II-C]{HY-I}, for general alphabets probability mass functions are read as densities with respect to a
common $\sigma$-finite reference measure and sums as integrals.

An $(n,M_1,\ldots,M_K,\varepsilon_n)$ code consists of codebooks $\{\bu_k(j):j\in[1:M_k]\}\subseteq\cX_k^n$, $k\in[1:K]$,
and mutually disjoint decoding regions $D_{\bj}\subseteq\cZ^n$, $\bj=(j_1,\ldots,j_K)\in\cM\equiv[1:M_1]\times\cdots\times[1:M_K]$,
with average error probability
\begin{equation}\label{eq:avgerr}
\varepsilon_n\equiv\frac{1}{M_{[1:K]}}\sum_{\bj\in\cM}W^n\bigl(D_{\bj}^c\st\bu_1(j_1),\ldots,\bu_K(j_K)\bigr),
\end{equation}
the messages being independent and uniformly distributed. Let $c_k:\cX_k^n\to[0,\infty)$ be cost functions and
$\bGamma=(\Gamma_1,\ldots,\Gamma_K)$ a cost constraint. The code satisfies the cost constraint if
$\frac1nc_k(\bu_k(j))\le\Gamma_k$ for all $k$ and $j$.

\begin{definition}\label{def:achievable}
A tuple $\bR=(R_1,\ldots,R_K)\in\R^K_+$ is \emph{$\varepsilon$-achievable} $(0\le\varepsilon<1)$ with cost constraint $\bGamma$ if
there exists a sequence of $(n,M_{1},\ldots,M_{K},\varepsilon_n)$ codes, $M_k=M_{k,n}$, satisfying the cost constraint with
\begin{equation}\label{eq:achievable}
\limsup_{n\to\infty}\varepsilon_n\le\varepsilon,\qquad\liminf_{n\to\infty}\frac1n\log M_{k,n}\ge R_k\quad(k\in[1:K]).
\end{equation}
The closure of the set of $\varepsilon$-achievable tuples is the $\varepsilon$-capacity region $C_{\bGamma}(\varepsilon\st\bW)$, and
$C_{\bGamma}(0\st\bW)$ is the capacity region. A $0$-achievable tuple is simply called achievable.
\end{definition}

A general input is a $K$-tuple $\bX=(\bX_1,\ldots,\bX_K)$ of sequences $\bX_k=\{X_k^n\}_{n=1}^\infty$, where $X_k^n$ takes values
in $\cX_k^n$ and $X_1^n,\ldots,X_K^n$ are mutually independent for each $n$; no consistency across $n$ is required. Let
$\cG_{\bGamma}$ be the set of general inputs with $\Pr\{\frac1nc_k(X_k^n)\le\Gamma_k\}=1$ for all $k$ and $n$ (the same notation as in Parts I and II). Given a general
input, let $Z^n$ be the output of $W^n$, so that
\begin{equation}
P_{X_1^n\cdots X_K^nZ^n}(\bx_1,\ldots,\bx_K,\bz)=\Bigl(\prod_{k=1}^KP_{X_k^n}(\bx_k)\Bigr)W^n(\bz\st\bx_1,\ldots,\bx_K),
\end{equation}
and write $X^n_{[1:K]}\equiv(X^n_1,\ldots,X^n_K)$. For every nonempty $\cS\subseteq[1:K]$, with complement
$\cS^c=[1:K]\setminus\cS$ and $X^n_\Sc=(X^n_k)_{k\in\Sc}$, define the normalized conditional information density
\begin{equation}\label{eq:density}
\iota^n_\cS\equiv\frac1n\log\frac{W^n(Z^n\st X^n_{[1:K]})}{P_{Z^n|X^n_\Sc}(Z^n\st X^n_\Sc)},
\end{equation}
where for $\cS=[1:K]$ the denominator is the unconditional density $P_{Z^n}(Z^n)$. For $K=2$, $\iota^n_{\{1\}}$,
$\iota^n_{\{2\}}$ and $\iota^n_{\{1,2\}}$ are the three densities of \cite[Eq.~(7)]{HY-I}. For a sequence of real random
variables $\{V^n\}$, the liminf and limsup in probability are \cite{HV93}, \cite{Han03}
\begin{equation}
\plimi V^n\equiv\sup\bigl\{\beta:\lim_{n\to\infty}\Pr\{V^n<\beta\}=0\bigr\},\qquad
\plims V^n\equiv\inf\bigl\{\alpha:\lim_{n\to\infty}\Pr\{V^n>\alpha\}=0\bigr\}.
\end{equation}
The \emph{conditional information spectral-inf} and \emph{spectral-sup rates} are
\begin{equation}\label{eq:spectral}
\Iinf(\bX_\cS;\bZ\st\bX_\Sc)\equiv\plimi\iota^n_\cS,\qquad\Isup(\bX_\cS;\bZ\st\bX_\Sc)\equiv\plims\iota^n_\cS,
\end{equation}
with the conditioning omitted for $\cS=[1:K]$. Obviously $\Iinf(\bX_\cS;\bZ\st\bX_\Sc)\le\Isup(\bX_\cS;\bZ\st\bX_\Sc)$.

\subsection{The Two Fundamental Lemmas}

The following two lemmas are the $K$-user versions of \cite[Lemmas~3 and 4]{Han98}. Their proofs are those of \cite{Han98}
written for $K$ users; they are given in Appendices~\ref{app:direct} and \ref{app:converse}, so that the paper is self-contained. The form of
Lemma~\ref{lem:direct} differs slightly from a literal extension of \cite[Lemma~3]{Han98} (Remark~\ref{rem:lemmas}).

\begin{lemma}[Direct fundamental lemma]\label{lem:direct}
Let $\bX=(\bX_1,\ldots,\bX_K)$ be a general input, let $M_1,\ldots,M_K$ be positive integers, and let
$\cA\equiv\{k:M_k\ge2\}$. Then, for every $n$ and every $\gamma>0$, there exists an $(n,M_1,\ldots,M_K,\varepsilon_n)$ code with
\begin{equation}\label{eq:direct}
\varepsilon_n\le\Pr\Bigl\{\bigcup_{\emptyset\ne\cS\subseteq\cA}\Bigl\{\iota^n_\cS\le\frac1n\log M_\cS+\gamma\Bigr\}\Bigr\}+(2^K-1)e^{-n\gamma}.
\end{equation}
If $\bX\in\cG_{\bGamma}$, the code can be chosen to satisfy the cost constraint $\bGamma$.
\end{lemma}

\begin{lemma}[Converse fundamental lemma]\label{lem:converse}
Every $(n,M_1,\ldots,M_K,\varepsilon_n)$ code satisfies, for every $\gamma>0$,
\begin{equation}\label{eq:converse}
\varepsilon_n\ge\Pr\Bigl\{\bigcup_{\emptyset\ne\cS\subseteq[1:K]}\Bigl\{\iota^n_\cS\le\frac1n\log M_\cS-\gamma\Bigr\}\Bigr\}-(2^K-1)e^{-n\gamma},
\end{equation}
where the densities $\iota^n_\cS$ are evaluated for the input in which $X^n_1,\ldots,X^n_K$ are independent and $X^n_k$ is uniformly
distributed on the codebook $\{\bu_k(j):j\in[1:M_k]\}$, i.e., $X^n_k=\bu_k(J_k)$ with $J_k$ uniform on $[1:M_k]$ (codewords are counted with
multiplicity; the same convention is used for the inputs induced by codes in Secs.~\ref{sec:general} and \ref{sec:sc}).
\end{lemma}

\begin{remark}[On the two lemmas]\label{rem:lemmas}
Lemmas~\ref{lem:direct} and \ref{lem:converse} are natural and straightforward extensions of \cite[Lemmas~3 and 4]{Han98},
as already noted in \cite[Sec.~X]{Han98}; their proofs are those of \cite[Appendix and Sec.~VI]{Han98} with the three
densities replaced by the $2^K-1$ densities \eqref{eq:density}. In Lemma~\ref{lem:direct}, codewords are drawn independently
according to $P_{X_k^n}$, and the decoder outputs the unique $\bj$ such that $\iota^n_\cS>\frac1n\log M_\cS+\gamma$ for all
nonempty $\cS\subseteq\cA$; a wrong message tuple differs from the transmitted one on a set $\cS\subseteq\cA$, and the expected
number of such tuples passing the test for $\cS$ is at most $e^{-n\gamma}$, which gives the term $(2^K-1)e^{-n\gamma}$. The
restriction to $\cS\subseteq\cA$ only reflects that no wrong message exists for a user with $M_k=1$, and Lemma~\ref{lem:direct}
coincides with the literal extension of \cite[Lemma~3]{Han98} when all $M_k\ge2$. In Lemma~\ref{lem:converse}, each of the
$2^K-1$ events contributes $e^{-n\gamma}$, exactly as the three events do in \cite[Appendix, Proof of Lemma~4]{Han98}. The
input in Lemma~\ref{lem:converse} belongs to $\cG_{\bGamma}$ whenever the code satisfies the cost constraint.
\end{remark}

\begin{remark}[The two lemmas as the starting point]\label{rem:starting}
Lemmas~\ref{lem:direct} and \ref{lem:converse} are information spectrum lemmas of complete generality: no assumption is
made on the channel or on the alphabets, and the inputs are arbitrary independent inputs. They are a tool that covers the
coding problems for the MAC as a whole. The coding theorems of the present paper are derived from them: the capacity region
of the general $K$-user MAC (Theorem~\ref{thm:general}) from both, and the strong converse for the $K$-user Gaussian MAC
(Theorem~\ref{thm:sc}) from Lemma~\ref{lem:converse} alone, combined with an evaluation of the information spectrum
(Lemma~\ref{lem:limsup}). Lemma~\ref{lem:limsup} itself is not a consequence of the two lemmas: it is a direct computation with
information densities, i.e., a part of the information spectrum calculus (cf.\ \cite[Remark~7]{HY-I}). The general formula for the
$\varepsilon$-capacity region \cite[Theorem~1]{HY-I},
\cite[Theorem~1]{HY-II}, which is the starting point of Parts I and II, and its $K$-user version \cite[Remark~13]{HY-I} are
derived from the same two lemmas in the same way as \cite[Theorem~5]{Han98}. In this sense the three parts of the series
rest on Lemmas~\ref{lem:direct} and \ref{lem:converse}, together with the information spectrum calculus for the channels at hand.
\end{remark}

%% ====================================================================
\section{Capacity Region of the General $K$-User MAC}\label{sec:general}

For a general input $\bX\in\cG_{\bGamma}$ let
\begin{equation}\label{eq:RW}
\cR_{\bW}(\bX)\equiv\Bigl\{\bR\in\R^K_+:R_\cS\le\Iinf(\bX_\cS;\bZ\st\bX_\Sc)\mbox{ for all nonempty }\cS\subseteq[1:K]\Bigr\}.
\end{equation}
For $K=2$, $\cR_{\bW}(\bX)$ is the region of \cite[Eqs.~(3.4)--(3.6)]{Han98}. The following is the $K$-user version of
\cite[Theorems~1 and 7]{Han98}; see also \cite[Corollary~1]{HY-I}.

\begin{theorem}[Capacity region of the general $K$-user MAC]\label{thm:general}
For a general $K$-user MAC $\bW$ with arbitrary alphabets and an arbitrary cost constraint $\bGamma$,
\begin{equation}\label{eq:general}
C_{\bGamma}(0\st\bW)=\bigcup_{\bX\in\cG_{\bGamma}}\cR_{\bW}(\bX).
\end{equation}
\end{theorem}

As noted in \cite[Remark~2]{Han98}, the right-hand side of \eqref{eq:general} need not be convex, so that time sharing is
not available for general MACs. The proof rests on the following lemma, which shows that no closure operation is needed in
\eqref{eq:general}; for $K=2$ this was noted in \cite[Remark~1]{Han98}.

\begin{lemma}[Closedness]\label{lem:closed}
The set $\cU_{\bGamma}\equiv\bigcup_{\bX\in\cG_{\bGamma}}\cR_{\bW}(\bX)$ is closed.
\end{lemma}

\begin{proof}
Let $\bR^{(m)}\in\cR_{\bW}(\bX^{(m)})$, $m=1,2,\ldots$, converge to $\bR$, where $\bX^{(m)}=\{X^{(m),n}_{[1:K]}\}_{n=1}^\infty\in\cG_{\bGamma}$.
Clearly $\bR\in\R^K_+$. Write $\iota^{(m),n}_\cS$ for the densities \eqref{eq:density} of $\bX^{(m)}$. Since
$R^{(m)}_\cS\le\Iinf(\bX^{(m)}_\cS;\bZ\st\bX^{(m)}_\Sc)$ for every nonempty $\cS$, we have
$\lim_{n\to\infty}\Pr\{\iota^{(m),n}_\cS\le R^{(m)}_\cS-1/m\}=0$ for every $\cS$. Hence there are integers $n_1<n_2<\cdots$ such that
\begin{equation}\label{eq:diag}
\Pr\bigl\{\iota^{(m),n}_\cS\le R^{(m)}_\cS-1/m\bigr\}\le1/m\quad\mbox{for all }n\ge n_m\mbox{ and all nonempty }\cS.
\end{equation}
Define the diagonal input $\bX$ by $X^n_{[1:K]}\equiv X^{(m),n}_{[1:K]}$ for $n_m\le n<n_{m+1}$ (and $X^n_{[1:K]}\equiv X^{(1),n}_{[1:K]}$ for $n<n_1$).
Then $\bX\in\cG_{\bGamma}$, because independence and the cost constraint are required for each $n$ separately, and, since the
density $\iota^n_\cS$ at blocklength $n$ depends only on $P_{X^n_{[1:K]}}$ and $W^n$, the densities of $\bX$ at
$n\in[n_m,n_{m+1})$ are those of $\bX^{(m)}$. Fix $\gamma>0$. There is $m_0$ such that $R_\cS-\gamma\le R^{(m)}_\cS-1/m$ for all
$m\ge m_0$ and all $\cS$. By \eqref{eq:diag}, $\Pr\{\iota^n_\cS\le R_\cS-\gamma\}\le1/m$ for $n_m\le n<n_{m+1}$, $m\ge m_0$, so that
$\Pr\{\iota^n_\cS\le R_\cS-\gamma\}\to0$ and $\Iinf(\bX_\cS;\bZ\st\bX_\Sc)\ge R_\cS-\gamma$. As $\gamma>0$ is arbitrary,
$\bR\in\cR_{\bW}(\bX)\subseteq\cU_{\bGamma}$.
\end{proof}

\begin{proof}[Proof of Theorem~\ref{thm:general}]
\emph{Direct part: $\cU_{\bGamma}\subseteq C_{\bGamma}(0\st\bW)$.} Let $\bX\in\cG_{\bGamma}$, $\bR\in\cR_{\bW}(\bX)$ and $\gamma>0$. Put
$\cA\equiv\{k:R_k>2\gamma\}$, $M_{k,n}\equiv\lceil e^{n(R_k-2\gamma)}\rceil$ for $k\in\cA$ and $M_{k,n}\equiv1$ for $k\notin\cA$.
Then $M_{k,n}\ge2$ exactly for $k\in\cA$, and, since $\lceil x\rceil\le2x$ for $x\ge1$, for every nonempty $\cS\subseteq\cA$,
\[
\frac1n\log M_{\cS,n}+\gamma\le R_\cS-2|\cS|\gamma+\frac{|\cS|\log2}{n}+\gamma\le R_\cS-\frac\gamma2\qquad\mbox{for }n\ge\frac{2K\log2}{\gamma}.
\]
Lemma~\ref{lem:direct} gives codes satisfying the cost constraint with
\[
\varepsilon_n\le\sum_{\emptyset\ne\cS\subseteq\cA}\Pr\Bigl\{\iota^n_\cS\le R_\cS-\frac\gamma2\Bigr\}+(2^K-1)e^{-n\gamma}.
\]
Since $R_\cS\le\Iinf(\bX_\cS;\bZ\st\bX_\Sc)$, each probability tends to zero by the definition of the liminf in probability.
Hence $\varepsilon_n\to0$, and the tuple $(\max\{R_k-2\gamma,0\})_{k=1}^K$ is achievable. Letting $\gamma\downarrow0$ and using the
closedness of $C_{\bGamma}(0\st\bW)$ gives $\bR\in C_{\bGamma}(0\st\bW)$.

\emph{Converse part: $C_{\bGamma}(0\st\bW)\subseteq\cU_{\bGamma}$.} Let $\bR$ be achievable, with a sequence of codes satisfying the cost
constraint, $\varepsilon_n\to0$ and $\liminf_n\frac1n\log M_{k,n}\ge R_k$. Let $\bX$ be the general input in which $X^n_k$ is uniform on
the $k$th codebook at blocklength $n$; then $\bX\in\cG_{\bGamma}$, and $\bX$ depends only on the sequence of codes. Fix $\gamma>0$.
For all sufficiently large $n$, $\frac1n\log M_{k,n}\ge R_k-\gamma$ for every $k$, hence $\frac1n\log M_{\cS,n}\ge R_\cS-K\gamma$ for
every $\cS$, and Lemma~\ref{lem:converse} gives, for every nonempty $\cS$,
\[
\Pr\bigl\{\iota^n_\cS\le R_\cS-(K+1)\gamma\bigr\}\le\varepsilon_n+(2^K-1)e^{-n\gamma}\to0.
\]
Thus $\Iinf(\bX_\cS;\bZ\st\bX_\Sc)\ge R_\cS-(K+1)\gamma$, and, $\gamma$ being arbitrary while $\bX$ does not depend on $\gamma$,
$\bR\in\cR_{\bW}(\bX)$. Hence the set of achievable tuples is contained in $\cU_{\bGamma}$, and so is its closure
$C_{\bGamma}(0\st\bW)$ by Lemma~\ref{lem:closed}.
\end{proof}

\begin{remark}[The $\varepsilon$-capacity region]\label{rem:eps}
In the same way as in the proof of \cite[Theorem~5]{Han98}, the two fundamental lemmas give the $K$-user general formula for the
$\varepsilon$-capacity region \cite[Remark~13]{HY-I}, in which, as explained in \cite[Remark~1]{HY-I}, rate tuples with negative
components are admitted before the closure is taken; for $K=2$ it is \cite[Theorem~1]{HY-I}. Only Theorem~\ref{thm:general} is needed below.
\end{remark}

%% ====================================================================
\section{The $K$-User Gaussian MAC: Achievability of the Polymatroid}\label{sec:gauss-direct}

From now on $\bW$ is the stationary memoryless $K$-user Gaussian MAC with $\cX_1=\cdots=\cX_K=\cZ=\R$,
\begin{equation}\label{eq:gauss}
Z^n=X^n_1+\cdots+X^n_K+V^n,\qquad V^n\sim\cN(0,NI_n),\quad N>0,
\end{equation}
where $V^n$ is independent of the inputs, so that
$W^n(\bz\st\bx_1,\ldots,\bx_K)=(2\pi N)^{-n/2}\exp(-\|\bz-\sum_k\bx_k\|^2/(2N))$. The cost functions are the powers
$c_k(\bx)=\|\bx\|^2$, and, as in \cite[Sec.~III]{HY-I}, we write the cost constraint as a power constraint
$\bP=(P_1,\ldots,P_K)$ with $P_k>0$. The set of inputs is still denoted $\cG_{\bGamma}$, now with $\bGamma=\bP$ (the same symbol is used in
Parts I and II). Thus $\cG_{\bGamma}$ is the set of general inputs with independent $X^n_1,\ldots,X^n_K$ and
$\|X^n_k\|^2\le nP_k$ almost surely; no assumption is made on the form of the distributions or on their means. For nonempty
$\cS\subseteq[1:K]$ let $P_\cS\equiv\sum_{k\in\cS}P_k$ and
\begin{equation}\label{eq:CS}
\sfC_\cS\equiv\frac12\log\Bigl(1+\frac{P_\cS}{N}\Bigr),\qquad
\Pi_{\bP}\equiv\bigl\{\bR\in\R^K_+:R_\cS\le\sfC_\cS\mbox{ for all nonempty }\cS\subseteq[1:K]\bigr\}.
\end{equation}
Since $\cS\mapsto\sfC_\cS$ is normalized, nondecreasing and submodular, $\Pi_{\bP}$ is a polymatroid \cite{TH98}
\cite[Remark~14]{HY-I}. It is the capacity region of the $K$-user Gaussian MAC, due to Cover \cite{Cov75} and Wyner \cite{Wyn74}
(cf.\ \cite[Remark~4]{HY-I}); in this paper it is recovered from Theorem~\ref{thm:general} (Corollary~\ref{cor:region}). The
densities \eqref{eq:density} exist for every input, since the conditional laws of $Z^n$ are Gaussian mixtures with positive
densities.

\begin{theorem}[Achievability of the polymatroid]\label{thm:gauss-direct}
For the $K$-user Gaussian MAC with power constraint $\bP$,
\begin{equation}\label{eq:gauss-direct}
\Pi_{\bP}\subseteq\bigcup_{\bX\in\cG_{\bGamma}}\cR_{\bW}(\bX)=C_{\bP}(0\st\bW).
\end{equation}
More precisely, for every $0<\rho<\min_kP_k$ there is $\bar\bX^\rho\in\cG_{\bGamma}$ with
\begin{equation}\label{eq:trunc-rate}
\Iinf(\bar\bX^\rho_\cS;\bZ\st\bar\bX^\rho_\Sc)\ge\sfC^\rho_\cS\equiv\frac12\log\Bigl(1+\frac{P_\cS-|\cS|\rho}{N}\Bigr)\quad\mbox{for all nonempty }\cS.
\end{equation}
\end{theorem}

\begin{proof}
\emph{Step 1 (Gaussian input and its spectrum).} Fix $0<\rho<\min_kP_k$, and let $X^n_1,\ldots,X^n_K$ be independent with i.i.d.\
components $X^{(n)}_{k,i}\sim\cN(0,P_k-\rho)$. For nonempty $\cS$, $U^n_\cS\equiv\sum_{k\in\cS}X^n_k+V^n$ is i.i.d.\
$\cN(0,N+P_\cS-|\cS|\rho)$ and independent of $X^n_\Sc$; hence $P_{Z^n|X^n_\Sc}(\bz\st\bx_\Sc)=\prod_{i=1}^ng_\cS(z_i-\sum_{k\in\Sc}x_{k,i})$, where
$g_\cS$ is the density of $\cN(0,N+P_\cS-|\cS|\rho)$, and
\[
\iota^n_\cS=\frac1n\sum_{i=1}^n\log\frac{g(V^{(n)}_i)}{g_\cS(U^{(n)}_{\cS,i})},
\]
with $g$ the density of $\cN(0,N)$. The summands are i.i.d.\ with finite variance and mean $\sfC^\rho_\cS$, so that $\iota^n_\cS\to\sfC^\rho_\cS$
in probability by the weak law of large numbers.

\emph{Step 2 (Truncation).} Let $A_k\equiv\{\bx\in\R^n:\|\bx\|^2\le nP_k\}$ and $\pi_k\equiv\Pr\{X^n_k\in A_k\}$; since $\E[(X^{(n)}_{k,i})^2]=P_k-\rho$ and $\mathrm{Var}[(X^{(n)}_{k,i})^2]=2(P_k-\rho)^2$, Chebyshev's inequality gives
$1-\pi_k=\Pr\{\frac1n\|X^n_k\|^2>P_k\}\le2(P_k-\rho)^2/(n\rho^2)\to0$. Let $\bar X^n_k$ have the density $P_{X^n_k}(\bx)\bm{1}[\bx\in A_k]/\pi_k$,
the $\bar X^n_k$ being independent, and let $\bar\bX^\rho$ be the resulting input; then $\bar\bX^\rho\in\cG_{\bGamma}$. Write $\bar Z^n$ and
$\bar\iota^n_\cS$ for its output and densities. For every $\bx_\Sc$ and $\bz$,
\[
P_{Z^n|X^n_\Sc}(\bz\st\bx_\Sc)=\int W^n(\bz\st\bx_{[1:K]})\prod_{k\in\cS}P_{X^n_k}(d\bx_k)\ge\int_{\prod_{k\in\cS}A_k}W^n(\bz\st\bx_{[1:K]})\prod_{k\in\cS}P_{X^n_k}(d\bx_k)
=\pi_\cS\,P_{\bar Z^n|\bar X^n_\Sc}(\bz\st\bx_\Sc),
\]
where $\pi_\cS\equiv\prod_{k\in\cS}\pi_k$. Hence, at every point of the support of $(\bar X^n_{[1:K]},\bar Z^n)$, the density
$\bar\iota^n_\cS$ is at least $\iota^n_\cS$ evaluated at the same point minus $\frac1n\log\frac1{\pi_\cS}$. Moreover, the joint density of
$(\bar X^n_{[1:K]},\bar Z^n)$ is at most $1/\pi_{[1:K]}$ times that of $(X^n_{[1:K]},Z^n)$. Consequently, for every $a\in\R$,
\begin{equation}\label{eq:trunc}
\Pr\{\bar\iota^n_\cS\le a\}\le\frac{1}{\pi_{[1:K]}}\Pr\Bigl\{\iota^n_\cS\le a+\frac1n\log\frac{1}{\pi_\cS}\Bigr\}.
\end{equation}
For $a<\sfC^\rho_\cS$, the right-hand side tends to zero by Step 1, since $\pi_k\to1$. This proves \eqref{eq:trunc-rate}.

\emph{Step 3 (Conclusion).} By \eqref{eq:trunc-rate}, $\Pi^\rho\equiv\{\bR\in\R^K_+:R_\cS\le\sfC^\rho_\cS\ \forall\cS\}\subseteq\cR_{\bW}(\bar\bX^\rho)$.
Let $\bR\in\Pi_{\bP}$ and $\theta_\rho\equiv\min_\cS\sfC^\rho_\cS/\sfC_\cS$, so that $\theta_\rho\to1$ as $\rho\downarrow0$. Then
$\theta_\rho R_\cS\le\theta_\rho\sfC_\cS\le\sfC^\rho_\cS$, i.e., $\theta_\rho\bR\in\Pi^\rho\subseteq\cU_{\bP}$. Since $\theta_\rho\bR\to\bR$ and $\cU_{\bP}$ is
closed (Lemma~\ref{lem:closed}), $\bR\in\cU_{\bP}$, and Theorem~\ref{thm:general} completes the proof.
\end{proof}

%% ====================================================================
\section{An Upper Bound on the Spectral-Sup Rates}\label{sec:limsup}

The converse direction rests on the following lemma, which is the main technical result of the paper.

\subsection{The Spectral-Sup Bound}

For $m\in[1:K]$ and $P>0$, let $a\equiv P+N$ (note that $a$ depends on $P$) and define the constants
\begin{equation}\label{eq:consts}
\delta_n(m,P)\equiv\frac{P}{2\sqrt n}\log\Bigl(1+\frac na\Bigr)+\frac{(m+1)P}{2n},\qquad
c(m,P)\equiv\frac{m(m-1)^2P/2+mNP+P^2/2}{a^2}.
\end{equation}
In Lemma~\ref{lem:limsup} and Remark~\ref{rem:quant} they are used with $m=|\cS|$ and $P=P_\cS$.

\begin{lemma}[Spectral-sup bound]\label{lem:limsup}
For the $K$-user Gaussian MAC with power constraint $\bP$, every input $\bX\in\cG_{\bGamma}$, every nonempty $\cS\subseteq[1:K]$, every $n$
and every $\gamma>0$,
\begin{equation}\label{eq:limsup-bound}
\Pr\Bigl\{\iota^n_\cS>\sfC_\cS+\delta_n(|\cS|,P_\cS)+5\gamma\Bigr\}\le\frac{c(|\cS|,P_\cS)}{\gamma^2\sqrt n}+\frac{(|\cS|+1)N}{2\gamma n}+e^{-n\gamma}.
\end{equation}
In particular, since the right-hand side does not depend on the input and $\delta_n(|\cS|,P_\cS)\to0$,
\begin{equation}\label{eq:limsup}
\Isup(\bX_\cS;\bZ\st\bX_\Sc)\le\sfC_\cS\quad\mbox{for every }\bX\in\cG_{\bGamma},\qquad\mbox{hence}\qquad
\sup_{\bX\in\cG_{\bGamma}}\Isup(\bX_\cS;\bZ\st\bX_\Sc)\le\sfC_\cS.
\end{equation}
\end{lemma}

\begin{remark}[The single-user case]\label{rem:single}
For $K=1$, Lemma~\ref{lem:limsup} states that the spectral-sup rate of the single-user Gaussian channel does not exceed
$\frac12\log(1+P/N)$ for any input satisfying the power constraint $P$, and Theorem~\ref{thm:sc} below reduces to the strong converse
for the Gaussian channel, for which an information spectrum proof was given in \cite[Theorem~3.7.4]{Han03} (see also \cite{Sha59}).
For $K=1$ no cross term appears in the proof below ($T_1=0$). For $K\ge2$, the cross terms $\Xi_{jk}$ between the inputs of different
users are the new element, and it is for them that the diffuse--exceptional decomposition of Step~1 is needed.
\end{remark}

\begin{remark}[The diffuse--exceptional decomposition]\label{rem:tan}
The proof below splits $\R^n$ into a high-dimensional \emph{diffuse} subspace, on which all users' second moments are small, and a
low-dimensional \emph{exceptional} subspace, on which the input energy may concentrate, and it uses a Gaussian reference law whose
variance is enlarged on the exceptional subspace. This idea is due to Tan \cite{Tan-cp}, \cite{Tan-sr}, who used it for exact second-order
converses for the two-user Gaussian MAC; as in \cite{Tan-sr}, we decompose the sum of all users' (second-moment) matrices. Since only a
first-order bound is needed here, the construction is much simpler: the threshold is $\sqrt n$, the variance on the exceptional subspace is
inflated by $n$, and all fluctuations are controlled by Chebyshev's inequality, without the Brascamp--Lieb inequality or any restriction of the
code. The present paper uses only this decomposition idea, not the results of \cite{Tan-cp}, \cite{Tan-sr}. Its conclusion is first-order only, but,
unlike the two-user second-order results of \cite{Tan-cp}, \cite{Tan-sr}, it holds for every number $K$ of users and for every rate tuple outside $\Pi_{\bP}$.
\end{remark}

\subsection{Proof of Lemma~\ref{lem:limsup}}

\paragraph*{Roadmap of the proof}
The proof bounds the unnormalized information density $i^n_\cS$ from above, realization by realization, by $n\sfC_\cS$ plus a few
random terms whose fluctuations are controlled by Chebyshev's inequality. After the reduction to the full set of users (Step~0), the
density is compared with a Gaussian reference density $q_n$ instead of the true output density (Steps~2 and 6; the change of measure
costs at most $\gamma$ with probability $1-e^{-n\gamma}$). If $q_n$ were simply $\cN(0,aI_n)$ with $a=P+N$, the comparison would fail for
inputs that put an energy of order $n$ into a few directions; this is why $\R^n$ is first split into a diffuse and an exceptional
subspace (Step~1, Remark~\ref{rem:tan}) and the variance of $q_n$ is inflated on the latter, at a rate cost $\delta_n\to0$. The remaining
random terms (Step~3) are a noise term, a cross term between the inputs of different users, and terms on the exceptional subspace;
the cross term is controlled by the independence of the inputs (Step~4, \eqref{eq:Xi}), and this is the only place where the
multiple-access structure enters (Remark~\ref{rem:indep}). Steps~5 and 6 collect the bounds.

Throughout the proof, $\cS$ and $\bX\in\cG_{\bGamma}$ are fixed, and $i^n_\cS\equiv n\iota^n_\cS$ denotes the unnormalized density.

\emph{Step 0 (Reduction to the full set of users).} Subtracting the inputs in $\Sc$ from $Z^n$ reduces $i^n_\cS$, realization by
realization, to the information density of an $|\cS|$-user Gaussian MAC for the full set of its users. Indeed, let
\[
\tilde Z^n\equiv Z^n-\sum_{k\in\Sc}X^n_k=\sum_{k\in\cS}X^n_k+V^n,\qquad\tilde W^n(\tilde\bz\st\bx_\cS)\equiv(2\pi N)^{-n/2}\exp\Bigl(-\frac{\|\tilde\bz-\sum_{k\in\cS}\bx_k\|^2}{2N}\Bigr).
\]
With $\bm s\equiv\sum_{k\in\Sc}\bx_k$, we have $W^n(\bz\st\bx_{[1:K]})=\tilde W^n(\bz-\bm s\st\bx_\cS)$. Since $X^n_\cS$, $X^n_\Sc$ and $V^n$
are mutually independent, conditioning on $X^n_\Sc=\bx_\Sc$ leaves the law of $\sum_{k\in\cS}X^n_k+V^n$ unchanged, so that
$P_{Z^n|X^n_\Sc}(\bz\st\bx_\Sc)=P_{\tilde Z^n}(\bz-\bm s)$. Hence, for every realization,
\begin{equation}\label{eq:reduction}
i^n_\cS=\log\frac{\tilde W^n(\tilde Z^n\st X^n_\cS)}{P_{\tilde Z^n}(\tilde Z^n)}.
\end{equation}
The right-hand side is the full-set information density of the $|\cS|$-user Gaussian MAC with inputs $X^n_k$, $k\in\cS$, which
satisfy the power constraints $P_k$. In the rest of the proof we relabel the elements of $\cS$ as $1,\ldots,m$ with $m\equiv|\cS|$, write
$Z^n$, $W^n$ and $i_n$ for $\tilde Z^n$, $\tilde W^n$ and $i^n_\cS$, and put
\begin{equation}\label{eq:shorthand}
P\equiv P_\cS=\sum_{k=1}^mP_k,\qquad a\equiv P+N,\qquad\sfC\equiv\sfC_\cS=\frac12\log\frac aN,\qquad X^n\equiv\sum_{k=1}^mX^n_k.
\end{equation}
For $m=1$ no cross term $\Xi_{jk}$ appears below; $T_1=0$ and the argument goes through unchanged.

\emph{Step 1 (Diffuse--exceptional decomposition).} Let
\begin{equation}\label{eq:M}
M_k\equiv\E\bigl[X^n_k(X^n_k)^\top\bigr],\qquad M\equiv\sum_{k=1}^mM_k,\qquad\tr M_k=\E\|X^n_k\|^2\le nP_k,\qquad\tr M\le nP.
\end{equation}
These matrices are positive semidefinite. Let $M=\sum_{i=1}^n\mu_ie_ie_i^\top$ be a spectral decomposition, and define the
exceptional and diffuse subspaces
\begin{equation}\label{eq:ED}
\cE_n\equiv\mathrm{span}\{e_i:\mu_i>\sqrt n\},\qquad\cD_n\equiv\cE_n^\perp,\qquad r_n\equiv\dim\cE_n\le\frac{\tr M}{\sqrt n}\le P\sqrt n,
\end{equation}
the bound on $r_n$ holding because the eigenvalues are nonnegative. Let $A_k\equiv\Pi_{\cD_n}M_k\Pi_{\cD_n}=\E[X^n_{k,\cD}(X^n_{k,\cD})^\top]$,
where $X^n_{k,\cD}\equiv\Pi_{\cD_n}X^n_k$. For every $v\in\R^n$, since $M-M_k\succeq0$ and $\Pi_{\cD_n}M\Pi_{\cD_n}=\sum_{\mu_i\le\sqrt n}\mu_ie_ie_i^\top$,
\[
v^\top A_kv=(\Pi_{\cD_n}v)^\top M_k(\Pi_{\cD_n}v)\le(\Pi_{\cD_n}v)^\top M(\Pi_{\cD_n}v)\le\sqrt n\,\|\Pi_{\cD_n}v\|^2\le\sqrt n\,\|v\|^2.
\]
Hence, since projections do not increase norms,
\begin{equation}\label{eq:Ak}
\lambda_{\max}(A_k)\le\sqrt n,\qquad\tr A_k=\E\|X^n_{k,\cD}\|^2\le nP_k\qquad(k\in[1:m]).
\end{equation}
The subspaces $\cE_n$, $\cD_n$ and the matrices $A_k$ depend only on the distribution of the input, not on its realization; this is
used in Step 6. Because $M$ is the sum over all users, \eqref{eq:Ak} holds for every user simultaneously.

\emph{Step 2 (Reference density).} Let $q_n$ be the density of $\cN(0,aI_n+n\Pi_{\cE_n})$, i.e., of the Gaussian law with variance $a$ on
$\cD_n$ and $a+n$ on $\cE_n$:
\begin{equation}\label{eq:qn}
q_n(\bz)=(2\pi a)^{-(n-r_n)/2}\bigl(2\pi(a+n)\bigr)^{-r_n/2}\exp\Bigl(-\frac{\|\bz_\cD\|^2}{2a}-\frac{\|\bz_\cE\|^2}{2(a+n)}\Bigr),
\end{equation}
with $\bz_\cD\equiv\Pi_{\cD_n}\bz$ and $\bz_\cE\equiv\Pi_{\cE_n}\bz$. The variance $a=P+N$ on $\cD_n$ is the output variance for independent
Gaussian inputs with full power, and the variance on $\cE_n$ is increased by $n$. On $\cE_n$ the input may put energy of order $n$ into
a few directions (by the power constraints, $\|\Pi(\bx_1+\cdots+\bx_m)\|^2\le mnP$ for every orthogonal projection $\Pi$), so that with variance $a$
there the term $\|\bz_\cE\|^2/(2a)$ in $-\log q_n$ could be of order $n$; with variance $a+n$ it is $O(1)$ (see \eqref{eq:b-exc}), at the price
$\frac{r_n}2\log(1+n/a)=O(\sqrt n\log n)$ in the normalization (see \eqref{eq:ell-id}). Let
\begin{equation}\label{eq:ell}
\ell_n\equiv\log\frac{W^n(Z^n\st X^n_{[1:m]})}{q_n(Z^n)}.
\end{equation}
Substituting $W^n$ and \eqref{eq:qn} and using $Z^n-X^n=V^n$, the constants combine as
\[
-\frac n2\log(2\pi N)+\frac{n-r_n}2\log(2\pi a)+\frac{r_n}2\log\bigl(2\pi(a+n)\bigr)=n\sfC+\frac{r_n}2\log\frac{a+n}a,
\]
so that, for every realization,
\begin{equation}\label{eq:ell-id}
\ell_n=n\sfC+\frac{r_n}2\log\Bigl(1+\frac na\Bigr)+\frac{\|Z^n_\cD\|^2}{2a}+\frac{\|Z^n_\cE\|^2}{2(a+n)}-\frac{\|V^n\|^2}{2N}.
\end{equation}

\emph{Step 3 (A pathwise upper bound).} We bound $\ell_n$ almost surely by $n\sfC$, a small deterministic term and four random terms.
Since $Z^n_\cD=\sum_kX^n_{k,\cD}+V^n_\cD$ and $X^n_\cD\equiv\sum_kX^n_{k,\cD}\in\cD_n$, we have $\langle X^n_\cD,V^n_\cD\rangle=\langle X^n_\cD,V^n\rangle$ and
\begin{equation}\label{eq:ZD}
\|Z^n_\cD\|^2=\sum_{k=1}^m\|X^n_{k,\cD}\|^2+2\sum_{1\le j<k\le m}\Xi_{jk}+\|V^n_\cD\|^2+2\langle X^n_\cD,V^n\rangle,\qquad\Xi_{jk}\equiv\langle X^n_{j,\cD},X^n_{k,\cD}\rangle.
\end{equation}
\emph{Power.} Since projections do not increase norms, almost surely
\begin{equation}\label{eq:b-power}
\sum_{k=1}^m\|X^n_{k,\cD}\|^2\le\sum_{k=1}^m\|X^n_k\|^2\le nP.
\end{equation}
\emph{Noise.} Since $\|V^n_\cD\|\le\|V^n\|$ and $\frac1{2a}-\frac1{2N}=-\frac{P}{2aN}$,
\begin{equation}\label{eq:b-noise}
\frac{\|V^n_\cD\|^2}{2a}-\frac{\|V^n\|^2}{2N}\le-\frac{P}{2aN}\|V^n\|^2=-\frac{nP}{2a}-\frac{P}{2aN}\bigl(\|V^n\|^2-nN\bigr).
\end{equation}
\emph{Exceptional component.} By the triangle inequality, $\|Z^n_\cE\|\le\sum_k\|X^n_k\|+\|V^n\|$, and by the Cauchy--Schwarz inequality
$(u_1+\cdots+u_{m+1})^2\le(m+1)(u_1^2+\cdots+u_{m+1}^2)$; using $n/(a+n)\le1$,
\begin{equation}\label{eq:b-exc}
\frac{\|Z^n_\cE\|^2}{2(a+n)}\le\frac{(m+1)\bigl(\sum_k\|X^n_k\|^2+\|V^n\|^2\bigr)}{2(a+n)}\le\frac{(m+1)P}{2}+\frac{(m+1)\|V^n\|^2}{2n}.
\end{equation}
Substitute \eqref{eq:ZD} into \eqref{eq:ell-id} and use \eqref{eq:b-power}--\eqref{eq:b-exc}. The term $+nP/(2a)$ from \eqref{eq:b-power}
cancels exactly the term $-nP/(2a)$ in \eqref{eq:b-noise}, and, almost surely,
\begin{equation}\label{eq:pathwise}
\ell_n\le n\sfC+n\hat\delta_n+T_1+T_2+T_3+T_4,
\end{equation}
where
\begin{multline}\label{eq:Ts}
n\hat\delta_n\equiv\frac{r_n}2\log\Bigl(1+\frac na\Bigr)+\frac{(m+1)P}2,\quad T_1\equiv\frac1a\sum_{j<k}\Xi_{jk},\quad T_2\equiv\frac{\langle X^n_\cD,V^n\rangle}a,\\
T_3\equiv-\frac{P(\|V^n\|^2-nN)}{2aN},\quad T_4\equiv\frac{(m+1)\|V^n\|^2}{2n}.
\end{multline}
By $r_n\le P\sqrt n$ in \eqref{eq:ED}, $\hat\delta_n\le\delta_n(m,P)$ of \eqref{eq:consts}.

\emph{Step 4 (Moments).} The second moments of $T_1,T_2,T_3$ are at most of order $n^{3/2}$, and $\E T_4$ is a constant. The independence of
the inputs is used only for $T_1$, and only pairwise independence is needed. Let $j\ne k$. Since $X^n_j$ and $X^n_k$ are independent, we
may first take the expectation over $X^n_{k,\cD}$ conditionally on $X^n_{j,\cD}$:
\begin{equation}\label{eq:Xi}
\E[\Xi_{jk}^2]=\E\bigl[(X^n_{j,\cD})^\top X^n_{k,\cD}(X^n_{k,\cD})^\top X^n_{j,\cD}\bigr]=\E\bigl[(X^n_{j,\cD})^\top A_kX^n_{j,\cD}\bigr]=\tr(A_kA_j).
\end{equation}
For positive semidefinite $A,B$, $\tr(AB)\le\lambda_{\max}(A)\tr B$, because $\tr((\lambda_{\max}(A)I-A)B)=\tr(B^{1/2}(\lambda_{\max}(A)I-A)B^{1/2})\ge0$.
With $A=A_j$, $B=A_k$ and \eqref{eq:Ak},
\begin{equation}\label{eq:Xi-bound}
\E[\Xi_{jk}^2]\le\lambda_{\max}(A_j)\tr A_k\le\sqrt n\cdot nP_k.
\end{equation}
There are $m(m-1)/2$ pairs. By the Cauchy--Schwarz inequality $(\sum_{j<k}\Xi_{jk})^2\le\frac{m(m-1)}2\sum_{j<k}\Xi_{jk}^2$, by \eqref{eq:Xi-bound},
and since each $k$ is paired with at most $m-1$ indices $j<k$, so that $\sum_{j<k}P_k\le(m-1)P$,
\begin{equation}\label{eq:T1}
\E[T_1^2]\le\frac{m(m-1)}{2a^2}\sum_{j<k}\E[\Xi_{jk}^2]\le\frac{m(m-1)^2P}{2a^2}\,n^{3/2}.
\end{equation}
No correlation between different pairs needs to be evaluated. For $T_2,T_3,T_4$ we condition on the inputs, of which $V^n$ is independent.
Given the inputs, $\langle X^n_\cD,V^n\rangle$ is Gaussian with mean $0$ and variance $N\|X^n_\cD\|^2$, and
$\|X^n_\cD\|^2\le(\sum_k\|X^n_k\|)^2\le m\sum_k\|X^n_k\|^2\le mnP$. Moreover, $\|V^n\|^2/N$ is chi-square with $n$ degrees of freedom, so
that $\mathrm{Var}\|V^n\|^2=2nN^2$ and $\E\|V^n\|^2=nN$. Hence
\begin{equation}\label{eq:T234}
\E[T_2^2]\le\frac{mNP}{a^2}\,n,\qquad\E[T_3^2]=\Bigl(\frac{P}{2aN}\Bigr)^2\cdot2nN^2=\frac{P^2}{2a^2}\,n,\qquad\E T_4=\frac{(m+1)N}2.
\end{equation}

\emph{Step 5 (Probability bound).} Fix $\gamma>0$. If $T_1+T_2+T_3+T_4>4n\gamma$, one of the four terms exceeds $n\gamma$. Chebyshev's inequality
$\Pr\{|T|>n\gamma\}\le\E[T^2]/(n\gamma)^2$ for $T_1,T_2,T_3$, Markov's inequality for $T_4\ge0$, the bounds \eqref{eq:T1} and \eqref{eq:T234},
$n\le n^{3/2}$, and \eqref{eq:pathwise} give
\begin{equation}\label{eq:ell-bound}
\Pr\Bigl\{\frac1n\ell_n>\sfC+\delta_n(m,P)+4\gamma\Bigr\}\le\sum_{t=1}^3\frac{\E[T_t^2]}{n^2\gamma^2}+\frac{\E T_4}{n\gamma}\le\frac{c(m,P)}{\gamma^2\sqrt n}+\frac{(m+1)N}{2\gamma n},
\end{equation}
for all $n$ and $\gamma>0$. The right-hand side and $\delta_n(m,P)$ depend only on $m$, $P$ and $N$, not on the input distribution.

\emph{Step 6 (From $q_n$ back to $P_{Z^n}$).} Write
\begin{equation}\label{eq:split}
i_n=\ell_n-\log\frac{P_{Z^n}(Z^n)}{q_n(Z^n)}.
\end{equation}
By the change-of-measure bound \cite[Lemma~1]{HY-I} with trivial conditioning, applied with $Q=q_n$ and $a=n\gamma$,
\begin{equation}\label{eq:com}
\Pr\Bigl\{\frac1n\log\frac{P_{Z^n}(Z^n)}{q_n(Z^n)}<-\gamma\Bigr\}\le e^{-n\gamma};
\end{equation}
indeed, the probability is $\int_{B}P_{Z^n}\le e^{-n\gamma}\int_Bq_n\le e^{-n\gamma}$ with $B\equiv\{\bz:P_{Z^n}(\bz)<e^{-n\gamma}q_n(\bz)\}$. This bound
requires $q_n$ to be fixed independently of the realizations of the input and of $Z^n$, which holds because $q_n$ was constructed in
Step~1 from the input distribution alone. If $\frac1ni_n>\sfC+\delta_n(m,P)+5\gamma$, then either $\frac1n\ell_n>\sfC+\delta_n(m,P)+4\gamma$ or the
event in \eqref{eq:com} occurs. Combining \eqref{eq:ell-bound}, \eqref{eq:split} and \eqref{eq:com}, and returning to the original
notation through \eqref{eq:reduction}, we obtain \eqref{eq:limsup-bound}. Finally, since $\delta_n(|\cS|,P_\cS)\to0$ and the right-hand side of
\eqref{eq:limsup-bound} tends to zero for every fixed $\gamma>0$, $\Pr\{\iota^n_\cS>\sfC_\cS+6\gamma\}\to0$ for every $\gamma>0$, which is
\eqref{eq:limsup}.\hfill$\Box$

\begin{remark}[What the proof uses]\label{rem:uses}
The proof uses only the power constraints, the independence of the inputs (pairwise, and only in \eqref{eq:Xi}), and the Gaussianity
and independence of the noise. Neither a particular form of the input distributions nor zero means are assumed; the means are absorbed
into the second-moment matrices $M_k$. The bound \eqref{eq:limsup-bound} is uniform over all inputs in $\cG_{\bGamma}$; this uniformity is not
needed for the strong converse (Theorem~\ref{thm:sc}), but it yields the quantitative statement of Remark~\ref{rem:quant}.
\end{remark}

\begin{remark}[The role of independence]\label{rem:indep}
The independence of the users' inputs cannot be dropped in Lemma~\ref{lem:limsup}. For $K=2$, let $X^n_1$ be the truncated Gaussian
input of Theorem~\ref{thm:gauss-direct} for user~1 and let $X^n_2\equiv\sqrt{P_2/P_1}\,X^n_1$, so that both power constraints hold but the
inputs are fully dependent. Then $X^n_1+X^n_2=(1+\sqrt{P_2/P_1})X^n_1$, and the same computation as in Theorem~\ref{thm:gauss-direct} shows
that $\Iinf(\bX_{\{1,2\}};\bZ)\ge\frac12\log\bigl(1+(1+\sqrt{P_2/P_1})^2(P_1-\rho)/N\bigr)$, which tends to
$\frac12\log(1+(\sqrt{P_1}+\sqrt{P_2})^2/N)>\sfC_{\{1,2\}}$ as $\rho\downarrow0$; this is the rate of the two users acting coherently. In the proof, the failure occurs exactly at \eqref{eq:Xi}: for these
inputs $\Xi_{12}$ is of order $n\sqrt{P_1P_2}$ almost surely, whereas for independent inputs \eqref{eq:Xi-bound} makes the root mean square of $T_1$
of order $n^{3/4}$. Thus the single moment computation \eqref{eq:Xi} is where the multiple-access structure enters.
\end{remark}

\subsection{Suprema of the Spectral Rates and the Capacity Region}

\begin{theorem}[Suprema of the spectral rates]\label{thm:sup}
For the $K$-user Gaussian MAC with power constraint $\bP$ and every nonempty $\cS\subseteq[1:K]$,
\begin{equation}\label{eq:sup}
\sup_{\bX\in\cG_{\bGamma}}\Iinf(\bX_\cS;\bZ\st\bX_\Sc)=\sfC_\cS=\sup_{\bX\in\cG_{\bGamma}}\Isup(\bX_\cS;\bZ\st\bX_\Sc).
\end{equation}
\end{theorem}

\begin{proof}
For every $\bX\in\cG_{\bGamma}$, $\Iinf(\bX_\cS;\bZ\st\bX_\Sc)\le\Isup(\bX_\cS;\bZ\st\bX_\Sc)\le\sfC_\cS$ by Lemma~\ref{lem:limsup}. Conversely, the truncated
Gaussian inputs of Theorem~\ref{thm:gauss-direct} give $\sup_{\bX}\Iinf(\bX_\cS;\bZ\st\bX_\Sc)\ge\sfC^\rho_\cS$ for every $0<\rho<\min_kP_k$, and
$\sfC^\rho_\cS\to\sfC_\cS$ as $\rho\downarrow0$. Hence
$\sfC_\cS\le\sup_{\bX}\Iinf(\bX_\cS;\bZ\st\bX_\Sc)\le\sup_{\bX}\Isup(\bX_\cS;\bZ\st\bX_\Sc)\le\sfC_\cS$.
\end{proof}

Let $\overline\cR_{\bW}(\bX)$ be defined as \eqref{eq:RW} with $\Iinf$ replaced by $\Isup$; for $K=2$ it is the region of \cite[Eqs.~(8.19)--(8.21)]{Han98}.

\begin{corollary}[Capacity region of the $K$-user Gaussian MAC]\label{cor:region}
For the $K$-user Gaussian MAC with power constraint $\bP$,
\begin{equation}\label{eq:region}
C_{\bP}(0\st\bW)=\bigcup_{\bX\in\cG_{\bGamma}}\cR_{\bW}(\bX)=\bigcup_{\bX\in\cG_{\bGamma}}\overline\cR_{\bW}(\bX)=\Pi_{\bP}.
\end{equation}
\end{corollary}

\begin{proof}
By Lemma~\ref{lem:limsup}, $\cR_{\bW}(\bX)\subseteq\overline\cR_{\bW}(\bX)\subseteq\Pi_{\bP}$ for every $\bX\in\cG_{\bGamma}$. Together with
Theorem~\ref{thm:gauss-direct}, $\Pi_{\bP}\subseteq\bigcup_{\bX}\cR_{\bW}(\bX)\subseteq\bigcup_{\bX}\overline\cR_{\bW}(\bX)\subseteq\Pi_{\bP}$.
Theorem~\ref{thm:general} identifies the first union with $C_{\bP}(0\st\bW)$.
\end{proof}

\begin{remark}[On Corollary~\ref{cor:region}]\label{rem:region}
The capacity region $C_{\bP}(0\st\bW)=\Pi_{\bP}$ of the $K$-user Gaussian MAC, due to Cover \cite{Cov75} and Wyner \cite{Wyn74}, is well
established. It is stated as such in textbooks and in many papers, for example by Cover and Thomas \cite{CT06}, El Gamal and Kim \cite{EGK11},
Tse and Viswanath \cite{TV05}, Tse and Hanly \cite{TH98}, and Shamai and Wyner \cite{SW97}. In the standard proofs, the direct part uses
Gaussian random codebooks with joint-typicality or successive-cancellation decoding, and the converse part is a weak converse based on
Fano's inequality, single-letterization through the memorylessness of the channel, and the maximum-entropy property of the Gaussian
distribution.

Corollary~\ref{cor:region} gives an information spectrum proof of this result. Both parts are derived from the general formula of
Theorem~\ref{thm:general}, and hence from the two fundamental lemmas. The direct part evaluates the spectral-inf rates of truncated
Gaussian inputs by the weak law of large numbers (Theorem~\ref{thm:gauss-direct}). The converse part uses the spectral-sup bound of
Lemma~\ref{lem:limsup}, which holds for every independent input satisfying the power constraints. Neither Fano's inequality nor
single-letterization is used. The second equality in \eqref{eq:region}, $\bigcup_{\bX}\cR_{\bW}(\bX)=\bigcup_{\bX}\overline\cR_{\bW}(\bX)$, is the
$K$-user version of the sufficient condition of Han \cite[Corollary~1]{Han98} for the strong converse property. The strong converse
itself is proved directly in the next section.
\end{remark}

%% ====================================================================
\section{The Strong Converse for the $K$-User Gaussian MAC}\label{sec:sc}

We use the notion of strong converse of \cite[Definition~2]{HY-I} (cf.\ \cite[Definition~7.12.1]{Han03}), in its $K$-user form.

\begin{definition}\label{def:sc}
A general $K$-user MAC $\bW$ with cost constraint $\bGamma$ has the \emph{strong converse property} if, for every
$\bR\in\R^K_+\setminus C_{\bGamma}(0\st\bW)$, every sequence of $(n,M_{1,n},\ldots,M_{K,n},\varepsilon_n)$ codes satisfying the cost constraint
and $\liminf_n\frac1n\log M_{k,n}\ge R_k$ $(k\in[1:K])$ has $\lim_{n\to\infty}\varepsilon_n=1$.
\end{definition}

\begin{theorem}[Strong converse]\label{thm:sc}
For every $K$, every $P_1,\ldots,P_K>0$ and every $N>0$, the $K$-user Gaussian MAC \eqref{eq:gauss} with noise power $N$ has the strong
converse property with power constraint $\bP$: every sequence of codes satisfying the power constraints whose rate tuple
$\bR\in\R^K_+$ lies outside $\Pi_{\bP}$ has $\varepsilon_n\to1$.
\end{theorem}

\begin{proof}
By Corollary~\ref{cor:region}, $\Pi_{\bP}=C_{\bP}(0\st\bW)$, so that the two statements of the theorem are the same; the argument below
uses only the second one and does not depend on Secs.~\ref{sec:general} and \ref{sec:gauss-direct}. Let $\bR\in\R^K_+\setminus\Pi_{\bP}$. There is a nonempty $\cS_0$ with
$\eta\equiv R_{\cS_0}-\sfC_{\cS_0}>0$. Take any sequence of codes satisfying the power constraints with $\liminf_n\frac1n\log M_{k,n}\ge R_k$
for all $k$, and let $\bX$ be the general input in which $X^n_k$ is uniform on the $k$th codebook at blocklength $n$. Then $\bX\in\cG_{\bGamma}$,
because every codeword satisfies the power constraint. Put $\gamma\equiv\eta/4$. For all sufficiently large $n$,
$\frac1n\log M_{\cS_0,n}\ge R_{\cS_0}-\gamma$, so that $\frac1n\log M_{\cS_0,n}-\gamma\ge\sfC_{\cS_0}+2\gamma$. Lemma~\ref{lem:converse}, keeping only the
event for $\cS_0$, gives
\begin{equation}\label{eq:sc-proof}
\varepsilon_n\ge\Pr\bigl\{\iota^n_{\cS_0}\le\sfC_{\cS_0}+2\gamma\bigr\}-(2^K-1)e^{-n\gamma}=1-\Pr\bigl\{\iota^n_{\cS_0}>\sfC_{\cS_0}+2\gamma\bigr\}-(2^K-1)e^{-n\gamma}.
\end{equation}
By Lemma~\ref{lem:limsup}, $\Isup(\bX_{\cS_0};\bZ\st\bX_{\cS_0^c})\le\sfC_{\cS_0}$, so that $\Pr\{\iota^n_{\cS_0}>\sfC_{\cS_0}+2\gamma\}\to0$. Hence
$\liminf_n\varepsilon_n\ge1$, i.e., $\varepsilon_n\to1$.
\end{proof}

In words: a rate tuple outside $\Pi_{\bP}$ asks some set $\cS_0$ of users to send jointly more than $\sfC_{\cS_0}$, the most that the channel can carry
about them when the inputs of the other users are known. The converse fundamental lemma then makes the code fail with the event for $\cS_0$
alone, and Lemma~\ref{lem:limsup} shows that this event has probability tending to one. The number of users enters only through the choice of
$\cS_0$, and, after Step~0 of the proof of Lemma~\ref{lem:limsup}, the users in $\cS_0$ form a Gaussian MAC with $|\cS_0|$ users and the full set as the only
set that matters.

The proof uses only Lemma~\ref{lem:converse} and Lemma~\ref{lem:limsup}, applied to the single input $\bX$ induced by the given sequence of
codes; the comparison with the existing proofs is given in Remark~\ref{rem:fongtan}. Combined with \cite[Remark~5]{HY-I}, Theorem~\ref{thm:sc}
gives the following result, which for $K=2$ is \cite[Eq.~(37)]{HY-I}.

\begin{corollary}[Fong and Tan {\cite[Theorem~1]{FT16}}: $\varepsilon$-capacity region]\label{cor:eps}
For the $K$-user Gaussian MAC with power constraint $\bP$, $C_{\bP}(\varepsilon\st\bW)=\Pi_{\bP}$ for every $0\le\varepsilon<1$.
\end{corollary}

\begin{proof}
$\Pi_{\bP}=C_{\bP}(0\st\bW)\subseteq C_{\bP}(\varepsilon\st\bW)$. If $C_{\bP}(\varepsilon\st\bW)$ contained a point outside the closed set $\Pi_{\bP}$, it would
contain an $\varepsilon$-achievable $\bR\notin\Pi_{\bP}$, and Theorem~\ref{thm:sc} would give $\varepsilon\ge\lim_n\varepsilon_n=1$, a contradiction.
\end{proof}

\begin{remark}[A per-blocklength form]\label{rem:quant}
The uniformity of \eqref{eq:limsup-bound} in the input gives a non-asymptotic statement. Combining \eqref{eq:converse} with
\eqref{eq:limsup-bound} exactly as in \eqref{eq:sc-proof}, with the constants $\delta_n(\cdot,\cdot)$ and $c(\cdot,\cdot)$ of \eqref{eq:consts} evaluated at
$m=|\cS|$ and $P=P_\cS$ (so that $a=P_\cS+N$), every $(n,M_1,\ldots,M_K,\varepsilon_n)$ code satisfying the power constraints with
$\frac1n\log M_\cS\ge\sfC_\cS+\delta_n(|\cS|,P_\cS)+6\gamma$ for some nonempty $\cS$ has
\begin{equation}\label{eq:quant}
1-\varepsilon_n\le\frac{c(|\cS|,P_\cS)}{\gamma^2\sqrt n}+\frac{(|\cS|+1)N}{2\gamma n}+2^Ke^{-n\gamma}.
\end{equation}
In particular $1-\varepsilon_n=O(n^{-1/2})$ for every fixed rate tuple outside $\Pi_{\bP}$. Taking $\gamma=\gamma_n$ with $\gamma_n\to0$ and
$\gamma_n^2\sqrt n\to\infty$, \eqref{eq:quant} shows that for every $0<\lambda<1$ there is $n_0(\lambda)$ such that every code with $n\ge n_0(\lambda)$
and $\varepsilon_n\le\lambda$ satisfies $\frac1n\log M_\cS\le\sfC_\cS+\delta_n(|\cS|,P_\cS)+6\gamma_n$ for all $\cS$. This is a per-blocklength bound of the
form described in \cite[Remark~5]{HY-I}, with a remainder of order slightly larger than $n^{-1/4}$. 
%This is considerably weaker than the
%remainders $O(\sqrt{(\log n)/n})$ of \cite{FT16} and $O(n^{-1/2})$ of \cite{Kos22}, and than the exact second-order results of
%\cite{Tan-cp}, \cite{Tan-sr} for two users: 
This is because the proof in this paper aims only at the first-order strong converse property required in Part I, and 
the rate $1-\varepsilon_n=O(n^{-1/2})$ is an artifact of Chebyshev's inequality. 
%and is presumably far from the true target: for discrete memoryless channels, for instance, the probability of correct
%decoding above capacity is known to decay exponentially \cite{Ari73}. No such exponential statement is claimed here.
\end{remark}

%% ====================================================================
\section{Discussion}\label{sec:discussion}

\begin{remark}[Relation to \cite{FT16} ]\label{rem:fongtan}
Fong and Tan \cite[Theorem~1]{FT16} proved the $\varepsilon$-strong converse for the Gaussian MAC with any number of users; since their bounds
hold at every blocklength, their proof yields the strong converse property of Definition~\ref{def:sc}, and the credit for Theorem~\ref{thm:sc}
goes to \cite{FT16} (cf.\ \cite[Remark~5]{HY-I}). Their proof quantizes the inputs, applies the wringing technique of Dueck \cite{Dueck81} and
Ahlswede \cite{Ahl82}, which is needed because expurgating message pairs destroys the independence of the codewords, and invokes the
hypothesis testing converse of Wang, Colbeck and Renner \cite{WCR09}.

The present proof is an elementary information spectrum proof specialized to the Gaussian MAC, and it needs no advanced technique. Lemma~\ref{lem:converse}
reduces the strong converse directly to the spectral-sup bound of Lemma~\ref{lem:limsup}, which is a computation with second moments and Gaussian
quadratic forms. No quantization is needed, and the detour through Ahlswede's wringing argument for discrete channels is avoided: since
Lemma~\ref{lem:converse} applies to the original code with independent uniform messages, nothing is expurgated, and the independence of the codewords
is used directly, in \eqref{eq:Xi} only. Nor is a hypothesis-testing meta-converse needed: Lemma~\ref{lem:converse} is proved by an elementary
argument using only the disjointness of the decoding regions and a change of measure (Appendix~\ref{app:converse}; cf.\ \cite[Appendix]{Han98}). (The Verd\'u--Han type bound can be
derived from the meta-converse of \cite{PPV10}, which is the tool of choice for second-order analysis, but the first-order strong converse does not
require it.) The price of this simplicity is that the proof is first-order only (Remark~\ref{rem:quant}) and relies on the Gaussian structure;
whether an analogous elementary bound holds for general discrete memoryless MACs, for which Part II uses Ahlswede's wringing-based bound
\cite[Lemmas~10 and 11]{HY-II}, is not addressed here.
\end{remark}

\begin{remark}[Role of Theorem~\ref{thm:sc} in Part I]\label{rem:role}
In Part I, Theorem~\ref{thm:sc} is applied to each component of a mixed (quasi-static fading) Gaussian MAC separately: a component whose
capacity region does not contain the rate tuple fails with probability tending to one, and, since the average error probability is the
$w$-average of the component error probabilities \cite[Eq.~(28)]{HY-I}, Fatou's lemma bounds $\liminf_n\varepsilon_n$ from below by the
outage probability. This gives the converse parts of \cite[Theorems~4, 6 and 7]{HY-I} for $0<\varepsilon<1$; for the $0$-capacity regions a weak
converse suffices \cite[Remark~6]{HY-I}.
\end{remark}

\begin{remark}[Strong converse versus $\varepsilon$-strong converse]\label{rem:distinction}
For a single channel, the strong converse property of Definition~\ref{def:sc} implies the $\varepsilon$-strong converse (Corollary~\ref{cor:eps}) but not
conversely \cite[Remark~3.5.1]{Han03}, and the distinction does not affect the capacity region. For mixed channels it is decisive: the
$\varepsilon$-strong converse of a component $\theta$ in outage gives only $\limsup_n\varepsilon_{\theta,n}=1$, which does not bound the $\liminf$ of
$\int\varepsilon_{\theta,n}\,dw(\theta)$, whereas the strong converse property gives $\lim_n\varepsilon_{\theta,n}=1$, to which Fatou's lemma applies
\cite[Remarks~6 and 7]{HY-I}, \cite[Theorem~6 and Remark~7]{HY-II}. In this sense the strong converse property is exactly the property of the
components that makes the $\varepsilon$-outage region the $\varepsilon$-capacity region of the mixed channel. Here it is obtained directly from
Lemmas~\ref{lem:converse} and \ref{lem:limsup}, and its per-blocklength form appears in Remark~\ref{rem:quant}.
\end{remark}

%% ====================================================================
\section{Concluding Remarks}\label{sec:conc}

Starting from two fundamental lemmas, the $K$-user versions of the random-coding and converse lemmas of Han \cite{Han98}, we derived the capacity region
of the general $K$-user MAC with cost constraint, showed that it contains the polymatroid capacity region of the $K$-user Gaussian MAC, proved that the
conditional information spectral-sup rates of the $K$-user Gaussian MAC never exceed $\frac12\log(1+P_\cS/N)$, and deduced that the suprema of the
spectral-inf and spectral-sup rates coincide. The strong converse for the $K$-user Gaussian MAC then followed directly from the converse fundamental
lemma, without wringing and without hypothesis-testing meta-converses. The central estimate is the spectral-sup bound of Lemma~\ref{lem:limsup}.

Together with Parts I and II, the series thus rests on the two fundamental lemmas, together with the information spectrum calculus for the
channels at hand: the general formulas used in Parts I and II are derived from them,
and the strong converse property needed in Part I for the mixed Gaussian MACs is derived here from the converse lemma. Two questions remain. First,
whether an elementary spectral-sup bound of the type of Lemma~\ref{lem:limsup} holds for other classes of MACs, such as general discrete memoryless MACs,
for which the strong converse is currently obtained through wringing. Second, whether the information spectrum argument can be sharpened to second order
\cite{Tan-cp}, \cite{Tan-sr}.

%% ====================================================================
\appendices
\section{Proof of Lemma~\ref{lem:direct}}\label{app:direct}

The proof is that of \cite[Lemma~3]{Han98}, written for $K$ users. Fix $n$, $\gamma>0$, a general input $\bX$ and $M_1,\ldots,M_K$. If
$\cA=\emptyset$, all $M_k=1$, the code with one message pair and decoding region $\cZ^n$ has $\varepsilon_n=0$, and \eqref{eq:direct} holds
trivially. Let $\cA\ne\emptyset$. For nonempty $\cS\subseteq\cA$ let
\[
\cT_\cS\equiv\Bigl\{(\bx_{[1:K]},\bz):\frac1n\log\frac{W^n(\bz\st\bx_{[1:K]})}{P_{Z^n|X^n_\Sc}(\bz\st\bx_\Sc)}>\frac1n\log M_\cS+\gamma\Bigr\},\qquad
\cT\equiv\bigcap_{\emptyset\ne\cS\subseteq\cA}\cT_\cS.
\]
\emph{Random code.} For each $k$, draw $\bu_k(1),\ldots,\bu_k(M_k)$ independently according to $P_{X^n_k}$, independently over $k$. For
$\bj\in\cM$ write $\bu(\bj)\equiv(\bu_1(j_1),\ldots,\bu_K(j_K))$. On observing $\bz$, the decoder outputs $\bj$ if $\bj$ is the unique tuple
with $(\bu(\bj),\bz)\in\cT$, and declares an error otherwise. If $\bX\in\cG_{\bGamma}$, every codeword satisfies the cost constraint
with probability one.

\emph{Error analysis.} By the symmetry of the random code, the error probability averaged over the random code is the same for every
transmitted tuple, and we may assume that $\bm{1}\equiv(1,\ldots,1)$ is sent; let $Z^n$ be the output. Then $(\bu(\bm1),Z^n)$ has the
law $P_{X^n_{[1:K]}Z^n}$, and an error occurs only if $(\bu(\bm1),Z^n)\notin\cT$ or $(\bu(\bj'),Z^n)\in\cT$ for some $\bj'\ne\bm1$. The
probability of the first event is the first term on the right-hand side of \eqref{eq:direct}.

Let $\bj'\ne\bm1$ and $\cS\equiv\{k:j'_k\ne1\}$. Since $j'_k\ne1$ is possible only if $M_k\ge2$, we have $\emptyset\ne\cS\subseteq\cA$, and
$\cT\subseteq\cT_\cS$. The codewords $\bu_\cS(\bj')$ are independent of $(\bu_\Sc(\bm1),Z^n)$, whose law is $P_{X^n_\Sc Z^n}$, and
$\bu_\Sc(\bj')=\bu_\Sc(\bm1)$. On $\cT_\cS$ we have $P_{Z^n|X^n_\Sc}(\bz\st\bx_\Sc)<e^{-n\gamma}W^n(\bz\st\bx_{[1:K]})/M_\cS$. Hence
\begin{align*}
\Pr\{(\bu(\bj'),Z^n)\in\cT_\cS\}&=\sum_{(\bx_{[1:K]},\bz)\in\cT_\cS}P_{X^n_\cS}(\bx_\cS)P_{X^n_\Sc}(\bx_\Sc)P_{Z^n|X^n_\Sc}(\bz\st\bx_\Sc)\\
&\le\frac{e^{-n\gamma}}{M_\cS}\sum_{\bx_{[1:K]},\bz}P_{X^n_{[1:K]}}(\bx_{[1:K]})W^n(\bz\st\bx_{[1:K]})=\frac{e^{-n\gamma}}{M_\cS},
\end{align*}
where $P_{X^n_\cS}=\prod_{k\in\cS}P_{X^n_k}$ and sums are read as integrals for general alphabets. For a given $\cS$ there are
$\prod_{k\in\cS}(M_k-1)<M_\cS$ tuples $\bj'$ with $\{k:j'_k\ne1\}=\cS$. By the union bound, the probability of the second event is at most
$\sum_{\emptyset\ne\cS\subseteq\cA}e^{-n\gamma}\le(2^K-1)e^{-n\gamma}$.

Consequently, the average error probability over messages and over the random code is at most the right-hand side of \eqref{eq:direct},
and there is a realization of the codebooks, with every codeword satisfying the cost constraint when $\bX\in\cG_{\bGamma}$, whose average
error probability does not exceed it.\hfill$\Box$

\section{Proof of Lemma~\ref{lem:converse}}\label{app:converse}

The proof is that of \cite[Lemma~4]{Han98}, written for $K$ users. Fix a code with codewords $\bu(\bj)\equiv(\bu_1(j_1),\ldots,\bu_K(j_K))$ and
decoding regions $D_{\bj}$, $\bj\in\cM$, and fix $\gamma>0$. Let $X^n_{[1:K]}$ be the input of the lemma, so that $X^n_{[1:K]}=\bu(\bJ)$ with
$\bJ=(J_1,\ldots,J_K)$ uniform on $\cM$, and let all output laws $P_{Z^n|X^n_\Sc}$, $P_{Z^n}$ be computed for this input. For nonempty $\cS$ and
$\bj\in\cM$ let
\[
B_\cS(\bj)\equiv\Bigl\{\bz:\frac1n\log\frac{W^n(\bz\st\bu(\bj))}{P_{Z^n|X^n_\Sc}(\bz\st\bu_\Sc(\bj_\Sc))}\le\frac1n\log M_\cS-\gamma\Bigr\},\qquad
B(\bj)\equiv\bigcup_{\emptyset\ne\cS\subseteq[1:K]}B_\cS(\bj),
\]
where $\bu_\Sc(\bj_\Sc)\equiv(\bu_k(j_k))_{k\in\Sc}$ and, for $\cS=[1:K]$, the denominator is $P_{Z^n}(\bz)$. Since $W^n(D^c_{\bj}\st\bu(\bj))\ge
W^n(B(\bj)\st\bu(\bj))-W^n(D_{\bj}\cap B(\bj)\st\bu(\bj))$ and $D_{\bj}\cap B(\bj)\subseteq\bigcup_\cS(D_{\bj}\cap B_\cS(\bj))$,
\begin{equation}\label{eq:conv-split}
\varepsilon_n\ge\frac1{M_{[1:K]}}\sum_{\bj\in\cM}W^n\bigl(B(\bj)\st\bu(\bj)\bigr)-\sum_{\emptyset\ne\cS\subseteq[1:K]}\frac1{M_{[1:K]}}\sum_{\bj\in\cM}W^n\bigl(D_{\bj}\cap B_\cS(\bj)\st\bu(\bj)\bigr).
\end{equation}
Fix a nonempty $\cS$. On $B_\cS(\bj)$ we have $W^n(\bz\st\bu(\bj))\le M_\cS e^{-n\gamma}P_{Z^n|X^n_\Sc}(\bz\st\bu_\Sc(\bj_\Sc))$. Writing $\bj=(\bj_\cS,\bj_\Sc)$
and using $M_{[1:K]}=M_\cS M_\Sc$,
\[
\frac1{M_{[1:K]}}\sum_{\bj\in\cM}W^n\bigl(D_{\bj}\cap B_\cS(\bj)\st\bu(\bj)\bigr)\le\frac{e^{-n\gamma}}{M_\Sc}\sum_{\bj_\Sc}\sum_{\bj_\cS}P_{Z^n|X^n_\Sc}\bigl(D_{(\bj_\cS,\bj_\Sc)}\st\bu_\Sc(\bj_\Sc)\bigr)\le e^{-n\gamma},
\]
because, for each fixed $\bj_\Sc$, the regions $D_{(\bj_\cS,\bj_\Sc)}$ with $\bj_\cS$ ranging over the messages of the users in $\cS$ are mutually disjoint and
$P_{Z^n|X^n_\Sc}(\cdot\st\bu_\Sc(\bj_\Sc))$ is a probability law (for $\cS=[1:K]$, the inner sum is $\sum_{\bj}P_{Z^n}(D_{\bj})\le1$). Finally,
$(\bu(\bJ),Z^n)$ has the law of $(X^n_{[1:K]},Z^n)$, and $Z^n\in B(\bJ)$ holds exactly when $\iota^n_\cS\le\frac1n\log M_\cS-\gamma$ for some nonempty $\cS$; hence the first
term of \eqref{eq:conv-split} is the probability on the right-hand side of \eqref{eq:converse}. Combining these with \eqref{eq:conv-split} gives \eqref{eq:converse}.
For general alphabets, sums over $\bz$ are integrals with respect to the reference measure.\hfill$\Box$

\section*{Acknowledgements}
\textit{Generative-AI use disclosure.} The authors used Anthropic's Claude (Opus 5.5) during the preparation of this manuscript. It assisted in drafting
the manuscript from the authors' instructions and proofs, including the organization of the sections, the presentation of the proofs of
Theorem~\ref{thm:general}, Lemma~\ref{lem:closed}, Theorem~\ref{thm:gauss-direct} and Appendices~\ref{app:direct} and \ref{app:converse}, the Remarks, and the editing of the English sentences. All
mathematical results and proofs have been checked by the authors, who take full responsibility for the content.

\end{document}